\documentclass[preprint,12pt,authoryear]{elsarticle}
\usepackage{amssymb}
\usepackage{amsmath}
\usepackage{longtable}
\usepackage{booktabs}   
\usepackage{array}        
\usepackage{geometry}     
\usepackage{multirow}   
\usepackage{float}
\usepackage{tabularx}
\usepackage{booktabs}
\usepackage{comment}

\journal{Journal of Rural Studies}

\begin{document}

\begin{frontmatter}



\title{Urban to Rural Migration in Eastern Europe: Unpacking digital ruralities through TikTok video analysis} 


\author{Anca-Simona Horvath} 

\affiliation {organization={Hong Kong University of Science and Technology (GZ)},department={Computational Media and Arts, @A: Art, Technology, Architecture}} 

\author{Cristian Toșa}
\affiliation{organization={Stavanger University}, department={Department of Safety, Economics and Planning}}
\author{Simai (Stella) Huang}
\affiliation {organization={Hong Kong University of Science and Technology (GZ)},department={Computational Media and Arts, @A: Art, Technology, Architecture}}

\begin{abstract}
Urban to rural migration is a less-researched phenomenon compared to its counterpart: rural to urban migration. In parts of Europe, an increasing number of people living in big urban centers within the country, or moving from other countries decide to relocate to rural areas. In this paper, we examine this phenomenon by analysing content posted on TikTok that documents this transition. We collected a corpus of 901 videos posted until late 2025, documenting urban to rural migration in Romania, under three hashtags, which have collectively been played a total of 24 million times at the time when we gathered the dataset. We analyse this corpus both quantitatively and qualitatively and discuss our findings through the lens of digital rurality - a theory based on Harvey's and Soja's spatial triad, applied to rural spaces, and based on the role of digital technologies as (re-)mediators of everyday lived experience. Specifically, we analyze the corpus as: (a) digital rural localities, (b) formal representations of the digital rural, and (c) everyday lives of the digital rural. We find that (a) Social media platforms enable new forms of paid labor that sometimes involve the commodification of the self in rural areas - although many of the creators we analyze do not explicitly acknowledge this with their audiences. (b) The digital rural gains new forms of representation, and rural areas in remote Romania are highly data-rich across not only TikTok but also Facebook and Instagram. (c) The everyday lives represented through the digital rural are sometimes idealized or romanticised. However, they serve as promoters for tourism and are used as sites to document and discuss a variety of topics including giving ample health advice, typically by non-specialists and sometimes criticizing Western medicine, expressing and promoting religious and political views but also acting as forms of general self-expression.

\end{abstract}

\begin{highlights}
\item urban to rural migration in Romania is studied by analyzing a corpus of TikTok videos
\item the motivations for people moving to rural areas, the mover profiles and experiences of the change are analyzed
\item the findings are discussed within the framework of digital rurality
\end{highlights}


\end{frontmatter}

\section{Introduction}

Rural areas in Eastern Europe are well-known to suffer from depopulation due to a variety of factors including an ageing population and both interior and exterior youth migration to larger urban centers or, primarily, to Western European countries. Youth migration is rooted in de-industrialization (which is a broader European phenomenon) and a lack of infrastructure, including schools, healthcare, services, or cultural facilities - but also transportation networks.

Romania is an exemplar of a country where rural depopulation brings forward a series of problems, most evident being the large urban-rural gap in terms of career opportunities and social mobility. But also - as recently made obvious during the country's 2024 canceled elections - stark ideological differences which play out in showing a divided society in political opinions. As a former member of the Eastern Bloc, Romania has a history of accelerated urbanization, industrialization, and rural to urban displacement, which at times was forced rather than voluntary, spanning the 1960s through the late 1980s \cite{Vais2022, Muntele2021}. After the fall of communism in 1989, the country underwent a long period of transition, where rural areas were depopulated even further, and experienced a period of decline. Starting 2007, when Romania joined the European Union, large scale migration of the working age population, from rural areas accelerated even more - but in this period, rural migrants prioritized Western European countries instead of Romanian urban centers \cite{PETRESCU2022}. 

A third period can be observed starting around 2019, coinciding with the Covid-19 pandemic. Since the start of Covid-19, when lockdowns forced digital work, but also imposed rigid conditions on public spaces, which were more strictly enforced in cities, a growing number of online communities started to document stories of urban-to-rural migration in Romania. Many of these stories are shared on the most popular social media platforms in country today, namely Facebook, TikTok and Instagram. While Romania's population is in decline overall, with urban areas losing 0.9\% of their population between 2024 and 2025 alone, and the country's population going from around 25 million in 1990 to 19 million in 2025, the last 5 years have seen a small increase in rural population of 0.04\% \cite{INS2026}. 

Analyzing urban phenomena through social media content is, by now, well-established, with work unpacking mobility \cite{ORSI2026, Hawelka2014}, social inequalities \cite{deSouza2025, deSouza2024, Morra2024}, and general use of urban public spaces and people's experiences in them \cite{Li2026, KANG2023, Ceccato2026, Horvath2024}. However, rural areas have been less researched by comparison and in that in particular of urban to rural migration as documented on social media media platforms such as TikTok.

Aiming to address this gap, in this paper, we ask: \textit{What are the main motivations for people to migrate from urban centers to rural areas in Romania today?} (RQ1) and \textit{What are the mover profiles - meaning is there a distinct demographics of people who make this move?} (RQ2) and finally: \textit{How do these movers experience the urban-to-rural lifestyle change?} (RQ3).

In the next section we discuss related work, including studies on urban-to-rural migration, as well as studies that analyze TikTok datasets to explore various phenomena. In Section 3, we present the materials and methods employed in conducting this study while Section 4 presents the findings, which are discussed in Section 5.
This study sheds light on various typologies of mover profiles, identifies key enablers and barriers influencing urban-to-rural migration and documents the phenomenon of digital rurality as it unfolds on one of the most popular social media platforms today. This can help shape actionable insights for rural planners and policymakers. 

\section{Related Work}

Social media platforms are reshaping how people relate to and engage with public spaces. As \cite{Tim2024} argue, social media can be understood as a reflection of on-the-ground processes and as an active force that shapes how the environment is perceived. Compared to other major social media platforms, many of which are older, TikTok content remains less researched than Twitter, Facebook, or Flickr \cite{Hart2022}. TikTok holds vast datasets of user-generated video content as well as data about how other users interact with this content. In Eastern Europe, urban-to-rural migration is becoming increasingly common, and short video platforms such as TikTok have become important channels for recording this social change \cite{Hautea2021}.

In this section, we first present other studies that have looked at TikTok content to analyse various phenomena, then discuss urban-to-rural migration and finally present the theoretical framework of digital rurality which we will employ to discuss the findings of our study.

\subsection{TikTok Content Analysis}

Existing TikTok content analysis research mostly adopted a mixed method approach, combining quantitative and qualitative techniques. Studies so far have looked at text or audio-visual content and various combinations of these. For example, prior research used machine learning algorithms to analyze user-generated text, such as video titles, tags, and comments with sentiment analysis tools such as  VADER (\cite{Hutto2014}) to measure public perceptions. \cite{Mulianingrum2025} demonstrated this method by using VADER to classify emotions in comments about fast fashion on TikTok and validated the results through manual coding. Other studies made use of multimodal analysis.  For example, \cite{Zha2026} proposed a scalable process that can extract and integrate features from text, video, and audio in more than 160,000 TikTok videos. In addition to a specific framework, \cite{Jorge2025} provide a comprehensive guide for TikTok content studies, including sampling strategy, user analysis, and post analysis, emphasizing the need to adapt traditional content analysis methods to short video platforms. 
Another strand of TikTok content analysis includes qualitative coding of video content and looking for specific themes such as medical information and political communication. \cite{Beach2025} analyzed user posted videos related to dermatological conditions linked to diabetes and illustrated how to systematically classify creators and evaluate the reliability of information. A number of studies have looked at medical advice posted on TikTok where for example \cite{Brittany2025} looked at mental health related content, and distinguished between creators with and without a medical background. Among studies that have looked at TikTok content relating specifically to rural areas, \cite{UNAYGAILHARD2023} investigated how farmers talk about climate and empathy. The authors find that the public's empathetic engagement with the farmer's climate dialogue remains at a shallow level of support through brief emotional reactions, while cognitive empathy that truly reflects deep understanding is relatively limited. \cite{HaoChen} used TikTok short videos and user comments to analyze the public's preferences for rural landscapes and found that users have a high overall evaluation of rural landscapes, but there are differences in preferences. 

\subsection{Urban to Rural Migration}
Recent reviews have revealed the diversity of urban to rural migration populations and the multidimensionality of motivations. \cite{Zhang2025} analyzed 58 cases from 1990 to 2024 and found that the impact of urban-to-rural migration is usually positive, as it promotes economic revitalization and ecological protection, while also noting the challenges of rural gentrification and conflicts between newcomers and locals. \cite{Erkan2025} reviewed 337 articles describing urban-to-rural migration as a heterogeneous group spanning different ages, educational backgrounds, and socioeconomic statuses, driven by both economic factors and non-economic motivations. 

For example, \cite{Takahashi2021} surveyed migrants arriving in Hokuto, Japan from 2017 to 2019, and developed an analytical framework to capture migration value. Further detailed analysis is conducted of the region's natural attributes, which are crucial to migrants' values in rural areas. However, as \cite{Carmen2022} note, the concept of "rural-urban interaction" has evolved, calling for new approaches that move beyond a static urban-centric lens and attend to the multi-functionality of rural spaces.

At the same time, there has been an emerging "digital trend" in urban-to-rural migration research that recognizes the role of social media in representing and shaping migration experiences. This trend is built on the academic foundation of earlier research on the relationship between social media and migration. Research by \cite{Lee2011} and \cite{Rianne2014} show how social media platforms can change migration networks, facilitate information exchange, and enable new forms of transnational belonging. Recently, studies have begun to apply these insights to the context of urban-to-rural migration, although most research so-far focuses on East Asia. An example based on TikTok's analysis comes from \cite{Li2025}, who used the data of Douyin (the Chinese version of TikTok) to test the institutional basis of trans provincial migration homesickness by looking at the thumbs-up emoji, showing how platform data reveal dimensions of migration sentiment. 

Despite these emerging contributions, the vast datasets of videos posted on TikTok remain under-explored in research. This is true also of videos that document urban-to-rural migration experiences. With migration and mobilities research, while TikTok content analysis typically focused on topics such as health (mis)information and political communication.

\subsection{Theoretical Framework}

In \cite{HE2026103958}, the term \textit{digital rurality} is introduced as a three-fold model which comprises: (a) digital rural localities, (b) formal representations of the digital rural, and (c) everyday lives of the digital rural. This model is based on the earlier work by \cite{HALFACREE2007125}, who took inspiration from Henry Lefevre's theory and to some extent Edward Soja's. In this view, there are spatial practices, or how ‘real’ space is perceived, there are representations of space - how various stakeholders conceptualize space (i.e. architects, planners, real estate developers), referring to space as conceived or ‘imagined’. And finally, there are spaces of representation which are associated with lived space, or \textit{"the spatial performance of everyday life"}. Based on the spatial triad, Halfacree proposes a spatial trialectic for rural space, which comprises: (a) rural localities inscribed through relatively distinctive spatial practices, linked to production and/or consumption activities; (b) formal representations of the rural and (c) the  everyday lives of the rural, which are inevitably subjective and diverse, and with varying levels of coherence \cite{HALFACREE2007125}. 

Based on these previous conceptualizations, the digital rural localities are understood as \textit{"socio-spatial changes resulting from the proliferation of digital technologies in rural areas"}. The formal representations of the digital rural are \textit{"digital transformations of rural spaces that are mobilized, stimulated, and promoted by power structures through specific representations of the digital rural"}.
Finally, the everyday lives of the digital rural are described as \textit{"the ways in which people interact with rural spaces through digital technologies in the course of their daily lives"} \cite{HE2026103958}.

This framework is useful in thinking about space in general including rural spaces. As virtual spaces become extensions of public and also private spaces, updating the spatial triad to include digital or virtual spaces is a natural progression. In this paper, we use the framework of digital rurality as a lens through which to discuss our findings. 

\section{Materials and Methods}
\begin{figure}
    \centering
    \includegraphics[width=0.9\linewidth]{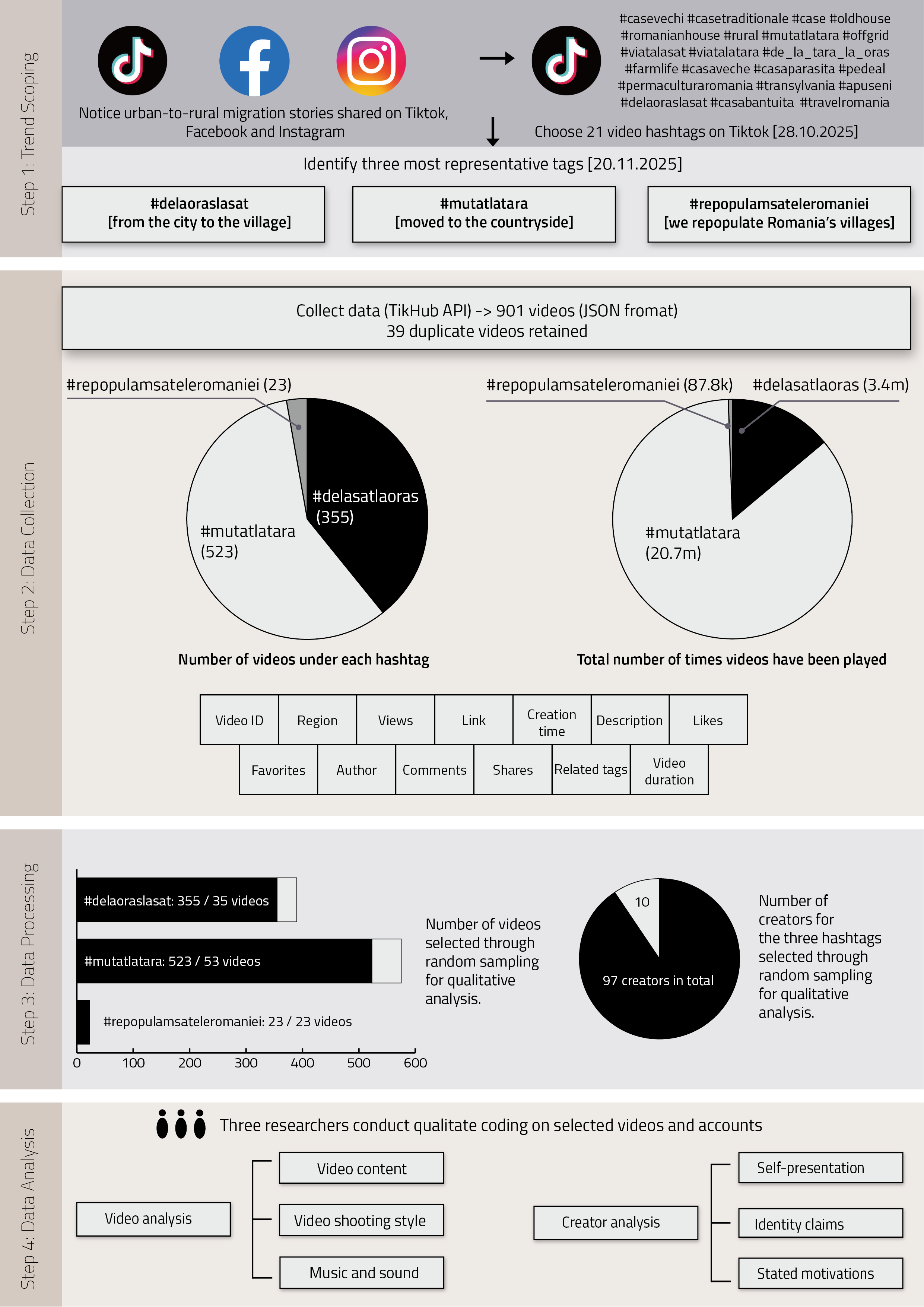}
    \caption{The four-step process in our methodology.}
    \label{fig1}
\end{figure}

We analyze a corpus of TikTok videos documenting urban to rural migration in Romania. We conduct both a quantitative and a qualitative  analysis on this corpus. Below, we describe the method for data collection and data analysis that includes selecting a series of hashtags that are relevant for our study, mining TikTok content via the TikTok API for all content that includes those hashtags, and subsequently analyzing this collected dataset (see Fig. \ref{fig1} which details ourentire research process).

\subsection{Data collection}
Two of the authors, both native speakers of Romanian, conducted a preliminary study on three social media platforms, namely Facebook, Instagram, and TikTok investigating the phenomenon of urban to rural migration in Romania, between September and October of 2025. We chose these three platforms because together they have a wide adoption in Romania today. Through a snowballing technique, we first identified a number of content creator profiles and Facebook groups where people share stories and document their urban-to-rural migration. Many of these groups are cross-platform, meaning, the same group with the same name would exist on both Facebook and Instagram, for example. Similarly, hashtags are also often cross-platform, so the hashtags we found on Facebook and Instagram also existed on TikTok. Moreover, many of the content creators have profiles on all these social media and in almost all cases, under the same names.

In this preliminary stage, 21 hashtags (show in Table \ref{tab:21-hashtags}) were identified, with the total number of videos and cumulative views recorded for each from October 28, 2025. From these 21 hashtags, we select three which we identified as most representative to adress our research questions and having most posts under them, namely: \textit{\#delaoraslasat} (from the city to the village), \textit{\#mutatlatara} (moved to the countryside), and \textit{\#repopulamsateleromaniei} ([we] repopulate Romania's villages).

Subsequently, one of the authors collected data for each of these three hashtags via TikTok's search function during November 2025. Once the data was collected, we filtered the videos and only retained those which had Romania or Ro under the 'Region' field. Among the three hashtags, \textit{\#mutatlatara} had the highest number of videos (523), which had been played 20.7 million times. \textit{\#delaoraslasat} followed with 355 videos and 3.5 million total views, while \textit{\#repopulamsateleromaniei} had the smallest count, with only 23 videos, but with 87.8k total views. This is detailed in Table \ref{tab:three-hashtags-1}.

We then collected all 901 videos through hashtag-based data crawling using the Tikhub API~\cite{TikHub}. We obtained data in JSON format of up to 20 videos at a time. For each video, we collected data as detailed in Table \ref{tab:video-data-fields}, including: creator, date when video was posted, number of times a video had been played until the date the data was collected. 39 of the collected videos were duplicates - meaning the same video had two or three of the hashtags we analyze. We kept these video duplicates as we otherwise would have to decide under which hashtag a specific video should be kept.

\begin{table}[H]
\centering
\caption{The 21 hashtags identified in the preliminary search conducted on Facebook, Instagram and TikTok}
\label{tab:21-hashtags}
\begin{tabular}{lr}
\toprule
\textbf{Hashtag} & \textbf{Translation/Description} \\
\midrule
\textit{\#farmlife} & Farming lifestyle \\
\textit{\#case} & Houses \\
\textit{\#rural} & Rural life \\
\textit{\#de\_la\_oras\_la\_tara}\ & From the city to the village \\
\textit{\#oldhouse} & Old houses \\
\textit{\#viatalatara} & Life in the countryside \\
\textit{\#transylvania} & Transylvania region \\
\textit{\#casabantuia }& Haunted house \\
\textit{\#pedeal} & On the hill \\
\textit{\#mutatlatara} & Moved to the village \\
\textit{\#casaveche} & Old house \\
\textit{\#casevechi} & Old houses \\
\textit{\#delaoraslasat} & From the city to the village \\
\textit{\#apuseni} & Apuseni Mountains \\
\textit{\#casetraditionale} & Traditional houses \\
\textit{\#romanianhouse} & Romanian house \\
\textit{\#casaparasiata} & Abandoned house \\
\textit{\#permaculturaromania} & Permaculture in Romania \\
\textit{\#offgrid} & Off-grid living \\
\textit{\#travelromania} & Travel in Romania \\
\textit{\#viatalasat} & Life in the village \\
\bottomrule
\end{tabular}
\end{table}

\begin{table}[H]
\centering
\caption{Data fields collected for each video}
\label{tab:video-data-fields}
\begin{tabular}{cc}
\toprule
\ Video ID, Description, Author, Region, Likes, Comments \\
\ Views, Shares, Favorites, Related Tags, Time Created, Duration, Link \\
\bottomrule
\end{tabular}
\end{table}


\begin{table}[H]
\centering
\caption{For the three selected hashtags - Number of creators, total number of videos of creators under a hashtag, and total number of followers for the creators as of November 10, 2025}
\label{tab:three-hashtags-1}
\begin{tabular}{lccc}
\toprule
\textbf{Hashtag} & \textbf{No. of creators} & \textbf{No. of Videos} & \textbf{No. of Followers} \\
\midrule
\textit{\#delaoraslasat} & 15 & 355 & 124,176 \\
\textit{\#mutatlatara} & 81 & 523 & 971,295 \\
\textit{\#repopulamsateleromaniei} & 1 & 23 & 16,323 \\
\bottomrule
\end{tabular}
\end{table}

\subsection{Data analysis}

Once the corpus was collected, we randomly sampled 10\% of videos under the first two hashtags. We analyzed: 52 videos under the \textit{\#mutatlatara} hashtag and 35 videos under the \textit{\#delaoraslasat} hashtag. Given the number of videos under the \textit{\#repopulamsateleromaniei} hashtag (n=23), adhering to the 10\% sampling rule would have resulted in a very small sample and so we included all 23 videos in our sample. Among the initially selected videos, there were six duplicate entries. 
The final sample size totaled 110 entries (Table \ref{tab:three-hashtags-2}). 

\begin{table}[htbp]
\centering
\caption{The selected videos for coding based on random sampling, for \textit{\#delaoraslasat} and \textit{\#mutatlatara} selected 10\%, for \textit{\#repopulamsateleromaniei}, we selected all 23 videos }
\label{tab:three-hashtags-2}
\begin{tabular}{lcc}
\toprule
\textbf{Tag} & \textbf{Translation} & \textbf{Videos} \\
\midrule
\textit{\#delaoraslasat} & From the city to the village & 35\\
\textit{\#mutatlatara} & Moved to the countryside & 52\\
\textit{\#repopulamsateleromaniei} & (We) repopulate Romania's villages & 23 \\
\midrule
\textbf{Total} & & \textbf{110}\\
\bottomrule
\end{tabular}
\end{table}

Next, we performed qualitative coding of the 110 videos. First, based on the research questions and a preliminary review of the sample content, we collaboratively developed an initial coding manual. Before analyzing the full sample, we conducted calibration exercises: three of the authors each independently coded five (15 videos in total) randomly selected videos from the \textit{\#delaoraslasat} dataset. Subsequently, we convened meetings to compare coding results, discuss discrepancies, and iteratively refine the manual definitions until consensus was reached. After calibration, the remaining videos were coded by three of the authors. Throughout this phase, we held regular consensus meetings to discuss our progress and impressions, to ensure we applied the coding criteria consistently and grounding emerging themes in collaborative interpretations. This analysis was conducted between November 2025 and February of 2026.

We analyze three aspects of the videos: (1) the video content (scenes, people, animals, activities, and objects); (2) video presentation (specifically referring to filming and editing techniques); and the (3) music and sound (including background music, sound effects, and narration). For videos which contained human narration (n=22), we also converted the sound into text for analysis.

Apart from the videos themselves, we also analyze more broadly profiles of the content creators.
For this, we conducted a targeted analysis of selected accounts. From the complete qualitative sample, we combined creators from two groups: the top 10\% from our qualitative sample and creators with the largest number of video outputs from the overall dataset. After removing duplicates, the final sample size is 13 (n=13). (Table \ref{tab:creators}). For these creators, we analyzed their self-presentation across multiple videos, focusing on their explicit or implicit identity claims, stated motivations, and their public personas. Two of the authors who are native speakers of Romanian, randomly selected 20 videos from each of the 13 creators and conductive a qualitative analysis of these videos.

\begin{longtable}
{p{0.12\columnwidth}p{0.12\columnwidth}p{0.12\columnwidth}p{0.12\columnwidth}p{0.45\columnwidth}}%

\label{tab:creators} \\

\toprule

\textbf{Creator} & \textbf{Videos from Sample} & \textbf{Videos from whole} & \textbf{Followers} & \textbf{Description} \\

\midrule
C1   & 24 & 64 & 16.2k & 30-40 year-old couple, bought house in countryside (Transylvania / Hunedoara - Târnava) \\
\hline
C2   & 13 &  113  &   44.9k    & 45-55 year-old woman who lived in Canada for 20 years, now living in a cătun in Romania, with her husband, a writer, partly promotes her own books in the videos \\
\hline
C3   & 5  &  55  &    58k   & 35-45 year old woman, psychotherapist, lived and worked in Germany for a number of years before moving to the Romanian countryside, posts about healthy living, including recipes for organic food, appears to be an influencer, lives in the Apuseni mountains (Transylvania) \\
\hline
C4   & 4  & 6   &    2   & 30-40 year old couple, digital nomads/former corporate, bought house in countryside, they post videos showing the renovation, both sharing equally the physical labour \\
\hline
C5   & 4  &  27  &    33.4k   & 30-35 year old woman, freelancer, moved to the countryside around 2022, shows videos about renovating the house and garden, but also shares political views and general life advice \\
\hline
C6   & 3  & 25   &   61.7k    & young male, documenting what he calls sustainable and self-sufficient living: built a house from scratch with recycled materials, most videos show details of the house (i.e. windows made from reused car doors); one of his videos showing a company describing a commercial wood and straw-based building solution presented at a fair;  \\
\hline
C7   & 3  &  3  &   25    & young/middle-aged couple who bought a new house in rural Romania. Asking for followers, but has only released three casual life short videos \\ 
\hline
C8   & 3  & 144   &   32k    & young (30 and 36) married couple, moved from Bucharest to countryside (Călărași county) in early 2025 to a house with 4000sqm garden; documenting their new everyday life in the countryside; they commute to Bucharest for their jobs \\
\hline
C9   & 3  &  9  &   5579    &  mid-aged male, documenting rural his everyday life around Timișoara \\
\hline
C10  & 3  &  149  &   2485   & young/mid-aged female, bought a farm in the countryside; shows her everyday life, explains different aspects of taking care of garden and animals \\
\hline
C11  & 0  &  24  &   15.9k  &  mid-aged male, documenting everyday life, part of it is in the countryside, however, part of his recent videos are of newsroll.ro - a self-declared news platform that showcases AI generated content in the form of news (both images and audio are AI generated) \\
\hline
C12  & 0  & 23   &  11.5k  & young/mid-aged couple, mainly sharing rural life about construction, bought house in the countryside \\
\hline
C13  & 0  & 22   &  4866  &  mid-aged male, showing his own reconstruction and reconditioning of Romanaian cultural built heritage (includes furniture, buildings in the countryside, etc); videos are slideshow images, sometimes showing before and after of reconditioning projects \\
\hline

\caption{The thirteen creators with the highest number of videos for the three hashtags as of December 2025, selected for qualitative analysis} 

\end{longtable}


\section{Findings}

In total, the videos under the three hashtags had been played more than 24.3 million times as of November 2025. This is significant given that (1) all three hashtags are in Romanian, (2) they all represent a relatively specific and narrow topic, and (3) the total population of Romania is slightly more than 19 million people (\cite{Eurostat2026}), with approximately 22 million speakers of the Romanian language around the world and roughly 8.5 million TikTok users in Romania (\cite{TikTok2026}). 
Moreover, 97 content creators produced videos under the three hashtags, who together have a total following of approximately 1.2 million TikTok users as of December 2025.

In the following subsections we first introduce the video content, as analyzed quantitatively using topic analysis applied to the descriptions of the videos we selected for qualitative analysis. Next, we synthesize findings under subsections that cover each of our three research questions: main motivations for people to migrate to rural areas (Section 4.2.), the mover profiles (Section 4.3.), and experiences of the change (Section 4.4.).

\subsection{Video content}

The content shared in videos under the three hashtags is highly diverse, but mostly focuses on: (a) documenting a variety of home and property renovation projects (both the process and the results), mostly done by the creators themselves. Notably, there is a good gender balance when it comes to engagement in construction work; (b) documenting the everyday life of taking care of livestock and caring for and planting vegetable gardens; (c) showing ways of cooking the wild plants people pick and vegetables they grow is also often presented - especially recipes to store food over the winter (i.e. pickles, jams etc); (d) Some of the videos are real estate advertisements promoting various properties in the Romanian countryside. (e) Finally, as some of the content creators have a broad following, some videos are creators talking directly to their followers on a variety of topics, including discussing religious and political views, and offering health advice (for example by using certain wild plants, creators describe various health benefits).

\begin{figure}
    \centering
    \includegraphics[width=0.9\linewidth]{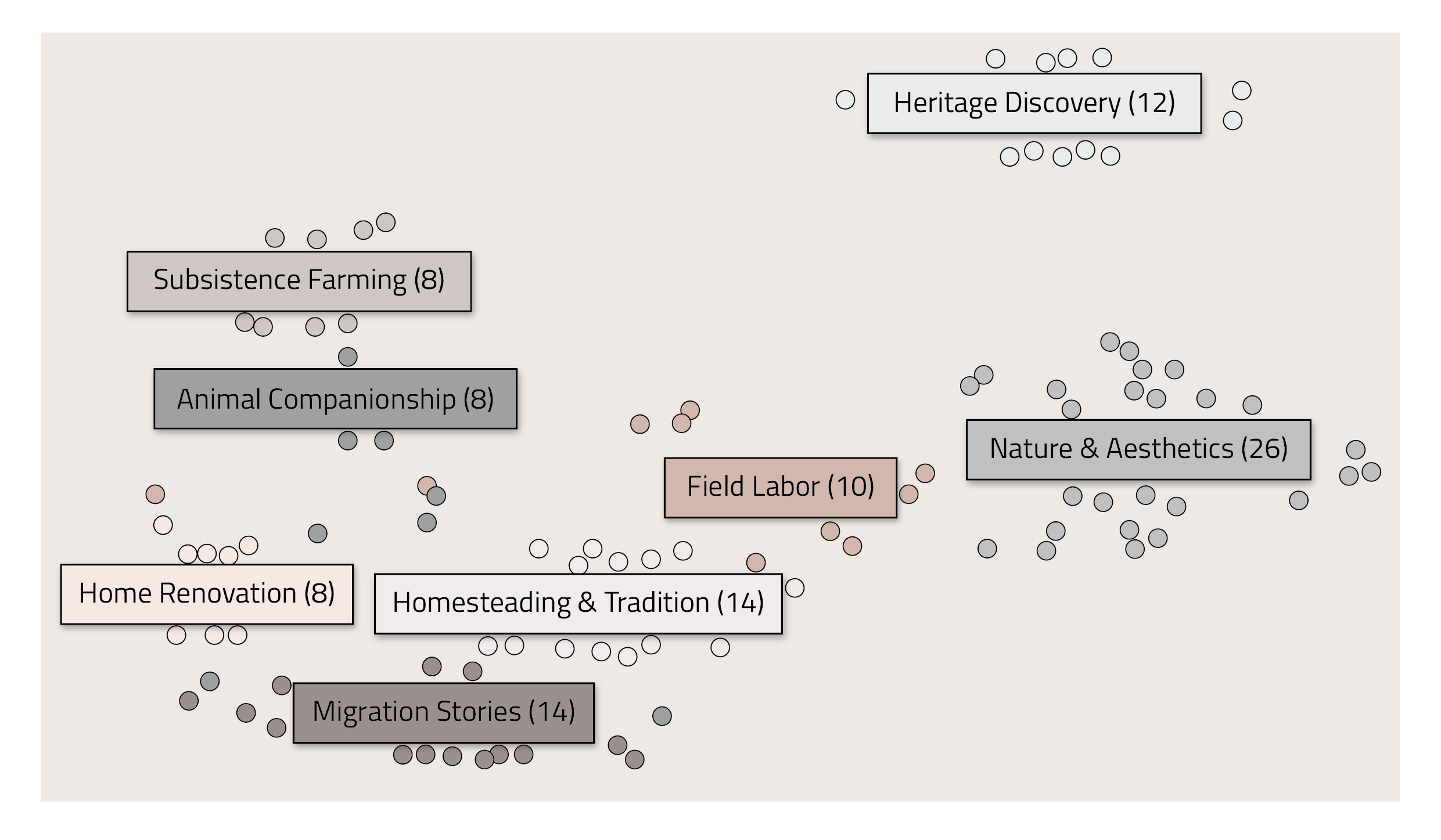}
    \caption{The mapping of rural migration topics.}
    \label{fig2}
\end{figure}

The topic modeling produced eight clusters (see Fig. \ref{fig2}), which map the thematic landscape of how urban-to-rural migration is narrated and performed across the three hashtags. The largest cluster, \textit{Nature \& Aesthetics} (n=26, 24.3\%), groups videos centered on the rural environment as a contemplative experience. \textit{Migration Stories} (n=14, 13.1\%) is testimonial in nature, with creators speaking directly to their audiences about the decision to relocate, experiences of returning from abroad, and (rarely) about challenges of rural adjustment. \textit{Homesteading \& Tradition} (n=14, 13.1\%) covers the practical and cultural dimensions of rural living, from identifying various species of plants and their possible uses in cooking and phytotherapy to property advice and traditional crafts. \textit{Heritage Discovery} (n=12, 11.2\%) is organised around the serialised documentary exploration of a rural property, presenting Romania's rural built environment as an object of discovery for an internationally oriented audience. \textit{Field Labor} (n=10, 9.3\%) documents the physical demands of agricultural and infrastructural work narrated with a tone of collective effort and pride. \textit{Subsistence Farming} (n=8, 7.5\%) covers the full productive cycle of rural food growing and animal husbandry, from foraging and harvest to preparation. \textit{Home Renovation} (n=8, 7.5\%) foregrounds property transformation as the central narrative, typically through before-and-after sequences that trace the progression of work across time. \textit{Animal Companionship} (n=8, 7.5\%) presents rural life through the lens of cohabitation with animals, using humor and sometimes 'cuteness' to attract views and likes.

The largest cluster, \textbf{Nature \& Aesthetics}, (n=26, 24.3\%) groups videos in which the rural environment is presented through scenery. Captions from creators function as invitations to contemplation: \textit{“A sunset in Cătunel”, “the walk in the morning”}, \textit{“Where do you look for the beauty?”}. The visuals are consistent across videos, and include close-ups of plants, animals, and seasonal transitions, accompanied by piano or calm acoustic music, and rarely featuring a voiceover, but in around half of the cases with a text layover. Without background voice, the videos center the landscape as the dominant presence. A subset of these videos make this logic explicit, pairing aesthetic content with quality-of-life claims: \textit{“For us, moving to the countryside was the best choice, from the effect it had on our mental health”}; \textit{“for a beautiful garden, you need to put in effort. But that does not overshadow the satisfaction.”} This aligns with the broader finding that creators present rural migration as a lifestyle aspiration, and this nature and aesthetic cluster is the most convincing means for that aspiration, reaching audiences through visual and auditory impact.

The \textbf{Migration Stories} cluster (n=14, 13.1\%) constitutes a comprehensive testimonial cluster. Creators narrate their own relocation as personal biography, addressing their audiences directly, typically in selfie mode with the landscape visible in the background, and open with community-building salutations, such as “hello, fellow Romanians,” before reflecting in their content video. Representative expressions include: \textit{“I returned from Canada to the top of the Apuseni Mountains. I built a house and a farm”}; \textit{“Did I have a hard time settling in on the mountaintop after returning from Canada?”}; \textit{“Three reasons why I left the city. Want more?”} These testimonials represent the enablers and barriers: diaspora return, the search for a slower pace, and the practical challenges of adjustment are all present in the verbal content. The selfie-mode filming convention, with creator in foreground, landscape behind, reinforces the testimonial logic by positioning the body in the countryside as the most credible form of evidence. With minimal or no post-production music, the creators are making the testimony as an unmediated confession.

\textbf{Homesteading \& Tradition} videos (n=14, 13.1\%) present rural living as a set of practices. Content ranges from phytotherapy demonstrations and plant identification labelled in Romanian, to traditional textile display and advice on purchasing rural property. An American-Romanian couple @far.farm.away, opens with \textit{“Join our progress as we take on a big fixer-upper in the countryside of Romania”} framing migration as an ongoing project with milestones and inviting their audience to take part in this journey. Another creator holds up her sleeve covered in stove charcoal with the caption, \textit{“Dedicated to all the beautiful village girls who stoke the fires”} explicitly constructing rural women's work as an identity rather than a burden. Filming in this cluster tends toward talking selfies for advice seekers, supplemented by close-up demonstrations, and accompanied by traditional instrumental music or relaxing ambient sound that indicates cultural belonging. This topic proves that movers approach rural transition pragmatically, combining affordability motivations with a deliberate construction of a new identity rooted in tradition.
The Heritage Discovery cluster (n=12, 11.2\%) is largely driven by a single account @unlocnumittarnava documenting the exploration and restoration of a rural property through multiple episodes: \textit{“… we bought a house in the countryside of Romania – Would you like to see more?”}; \textit{“Do you want to see what we found in the attic?”}; \textit{“Old objects once had so many personalities.”} Crucially, the use of AI-generated English voiceover, a consistent marker across this cluster, signals a distinct outward orientation toward an international audience rather than toward the Romanian diaspora community, as in the Migration Stories cluster. This linguistic split points to the co-existence of at least two publics for rural migration content: an internal Romanian audience and a global one, focusing on destination and discovery. Background music, such as relaxing instrumental, flute, atmospheric sound effects, varies, while the serialised episodes convert migration into a progressive narrative.

\textbf{Field Labor} videos (n=10, 9.3\%) move between collective achievement and practical instruction, documenting agricultural and infrastructural work. Content descriptions cover fencing, potato planting, vine rejuvenation, terrace construction, and wood chopping; captions blend pride with audience address: \textit{“We have made some progress with the work on the land: we have finished the fence”}; \textit{“Work on expanding the future terrace is progressing slowly but surely. Let's rejoice together in the progress”}; \textit{“how did I make the fence for the garden?”} The dominant filming technique is fast-forward or time-lapse, set to popular or upbeat music, which makes demanding work appear satisfying and manageable. A counter-register also appears within the cluster: a small number of videos use only ambient sound and the creator's unmediated voice, giving a more unguarded sense of rural labour as genuinely hard but personally meaningful.

The \textbf{Subsistence Farming} cluster (n=8, 7.5\%) documents the full productive cycle of rural food growing, foraging, and animal husbandry, from garden harvest to prepared dish. Content covers duck and chicken care, chick hatching, Easter egg dyeing, mushroom foraging-to-pasta sequences, and omelette preparation from garden vegetables. Captions reflect a farm-to-table culture: \textit{“Simple food without too many complications with what we have in the garden”}; \textit{“Our little darlings, our […] chicks, the first chicks born in our home.”} In terms of background music, several videos use traditional Romanian folk music and manele, while others use international pop music. This positions Subsistence Farming as a reference for those who share the cultural reference, grounding the content firmly in a local culture, and also for those who distinguish self-sufficient rural life from cosmopolitan lifestyles.

\textbf{Home Renovation} videos (n=8, 7.5\%) foreground property transformation as the central narrative of rural migration, and are organised around the before-and-after logic of property transformation. The dominant form is the photo slideshow: \textit{“Day 1 of our adventure”}; \textit{“We started the visits – this is not how we bought the house, but we got to this stage after many hammer blows”}; \textit{“Together we can do it.”} Several videos also make use of AI-generated English voiceover, again addressing an international audience rather than a Romanian one. Background music across the cluster tends toward popular or American pop, aligning renovation with aspirational lifestyle content.

The \textbf{Animal Companionship} cluster (n=8, 7.5\%) includes videos presenting animals as companions (often described as cute, and friendly) rather than resources. Videos include quotes such as: \textit{“We never get bored”}; \textit{“I trained the chickens”}; \textit{“Meet Alba, our new best friend!”}. Often, filming is first-person handheld in selfie mode, with no post-production music in most of the cases, and only including the creator's own voice and ambient farm sounds. Where music is present it is light and popular.

\subsection{Main motivations for people to migrate from urban centers to rural areas}

To a certain degree, almost all analyzed mover profiles are in search of what is perceived as self-sufficient lifestyles that would not be easily achieved by living in a city. For example, in one video, C2 in selfie mode, sitting in a field of flowers, makes a summary of her life and how she returned home to Romania after living in Canada for 20 years: \textit{"Hello brother Romanians, I am happy and free, in connection with good God, I left at some point far away to look for freedom, happiness, but all I could find was an empty space that slaughtered my soul. This brought me back to this wonderful land, where I can say ... God helps, that I am happy. Diary entrance - may God help!"}

Many of the creators choose to move to rural areas because of housing prices in large urban centers such as Bucharest or Cluj - so this migration is fueled by the housing affordability crisis on the one hand. For example, C8 has a video describing how they as a coupe made the decision to move to the countryside, and explaining that they sold their apartment in Bucharest and decided to buy an old house where they could move in right away. Their videos often talk about the cost of various renovation projects around the house and garden.

Another reason often stated is being in search of \textbf{\textit{a slower pace for living}} and a healthier life in general. In one of C5's videos, showing a renovated historical house, and herself coming into the frame feeding cats she details her views on the future. A dog - Romanian Shepherd - is also part of the frame. The video has text overlay stating: \textit{Hi, I am Alexandra and these are my predictions three years after moving to the countryside: (1) time humans spend in nature will become luxury, (2) social media decline and returning to physical events, (3) minimalism, de-growth and consumerism becoming saturated, (4) more and more people will move to the countryside, (5) more and more people will look for peace as opposed to social illusions}. Another example is
C7, who, in one of his videos, describes that he would do an experiment of getting off social media, and using a non-smart phone for a week. He explains experiencing a certain technological malaise, presents a pertinent critique of surveillance capitalism, and wanting to spend fewer hours during the day consuming commercials and being targeted by marketing campaigns. He advocates for a slower living. In a follow-up video, he states he managed to only use a non-smart phone for two weeks, and describes ways in which this has helped him: connect with other people (i.e. asking for directions rather than using a maps tool), it has reduced his dependence on new technologies, and has helped him to reconnect with other people. In his view, this made him more self-reliant.
This search stronger connection to nature, perceived as being more healthy, both physically and mentally is connected also to \textit{technolgical malaise}, which many creators mention in their videos (i.e. C2, C3, C5).

Apart from this, being able to work flexibly as digital nomads especially as facilitated during the Covid-19 pandemic appears as another reason to spark the decision o moving.

In summary, enablers for people to decide to move to the countryside include: (a) lower cost of property, including for some already owning a house in the countryside; and (b) the Covid-19 pandemic which has made people reconsider whether living in cities is indeed the most attractive lifestyle, but has also facilitated digital work; (c) searching for a slower pace for life and what is perceived as a more sustainable, healthier life (mentally and physically), and, to some extent, technological malaise.

\subsection{Mover profiles}

The mover profiles are diverse. Two of the creators returned to Romania after long periods lived abroad (C2 and C3), or are internationals who decide to move to the Romanian countryside (C1). The rest move from cities such as Bucharest, Cluj and Timișoara to villages across the country, but primarily in Transylvania, and around Bucharest. A number the creators are young to middle aged couples (C1, C4, C8, C12). Some are 30-40 year old males documenting building or renovation, and sustainable living (i.e. C6, C9, C11, C13) and others are 30-55 year old females documenting everyday life, but also often offering medical opinions and health advice. 

For example, C3 is a 35-45 year-old woman who makes visually coherent videos (in her own distinctive style) showing her farm with animals (including a number of Highland cows, dogs, cats), cooking videos and everyday life on the farm. Sometimes, her daughter (5-8 year old) appears in videos as well, either as a narrator or as a character. The videos include mixed messages. For example, one of the videos shows a few Highland cows. The videos is narrated by the daughter (a child's voice) who tells the audience about the cows, their ages and their names. Initially, it appears like the video is promoting the farm of the creator as turistic destination. However it also includes vaguely nationalistic references -  unclear in purpose (meaning they appear out of context): \textit{"This is Romania, a country that I love, a country that I long for" [...] "May God give us good leaders"}. C3's videos also include medical advice, partly discussing problems of Western medicine, and promoting healthy natural alternatives using plants found in the landscapes of Trasylvania. Similarly, C2 introduces various local wild plants and talks about how they could be used to treat depression, as replacements for more common drugs. C2 also appears somewhat critical of Western medicine. C5 has a video where she follows up on one of her previous posts about how a cream recommended by her dermatologist burned her face, and how she is currently using another cream made of natural materials which works much better for her. In the follow-up video, she explains that she does believe in science, and she will not that she will not only recommend traditional solutions, and she will not build on nostalgia for the past, but that instead she does believe in modern science, discoveries and progress.

In one of their videos, C8 describe the process they undertook in moving from Bucharest to the countryside. They explain what they looked for in buying a houses whether the village had a school, a kindergarden, medical facility and pharmacy. Apart from that, she mentions checking that the legal status of the property is in order. They sold their apartment in Bucharest, and bought a house in a village. They preferred renovating the house rather than building a new one from scratch for financial reasons, but also because they needed a place they could move to right away.

\subsection{Experiencing the urban-to-rural lifestyle change}

Planting and caring for gardens, feeding animals, and general renovation and maintenance of properties, but also cooking and storing food for the winter are all part of the new everyday lived experience. Videos document this new life in ample detail, and for the most part, the change is presented as positive, a way to reconnect with nature, ones self, spend less time online.

People sometimes experience difficulties in their transitions to new lifestyles: they discover that rural life requires hard physical work and new routines (closer to natural cycles), to which most are not used (i.e. C5 has some videos showing herself having a hard time waking up when a construction worker comes to fix her roof). C2 describes in one of her videos the things she finds hard about her new life, the main one being having to keep animals as livestock rather than as companions. She talks about how she could never get over having to slaughter animals that she cared for over a period of time. Similarly, C8 states that the reason they have birds around the garden is for eggs, and that the birds will die of old age, and not by slaughter. In one of the videos, C8 confesses that winters are hard in rural areas especially because the roads connecting the village to the capital city do not have asphalt everywhere. One AI generated videos under newsroll.ro in the profile of C11 presents (in a format that resembles news) the Romanian countryside as having extremely poor infrastructures (i.e. only dirt roads, many villages not being connected to water and gas). Another C11 newsroll.ro video, discusses social issues related to parents leaving the country to work abroad, and leaving their children behind to be cared for by the grandparents (this is a well-known phenomenon in Romania's rural areas). However, these videos are exceptions. While many videos show the creators conducting physical labor, and them sometimes mentioning that they find this hard, for almost all videos the lifestyle in the countryside is presented as desirable, quiet, a place for people to reconnect with themselves, and nature. There are many more videos that romanticize the countryside, than those showing unpleasant aspects of rural life.

\section{Discussion}
Our findings address the three research questions as follows. With respect to motivations (RQ1), urban-to-rural migration in Romania is driven by a combination of housing affordability pressures, a search for slower and more self-sufficient lifestyles, and the conditions created by the Covid-19 pandemic — which both normalised remote work and prompted a broader reassessment of urban living. With respect to mover profiles (RQ2), creators are predominantly young to middle-aged, digitally literate, and include both domestic urban dwellers and members of the Romanian diaspora returning from Western Europe or North America. With respect to lived experience (RQ3), the transition is represented on TikTok as largely positive and aspirational, centred on reconnection with nature, self-sufficiency, and identity reconstruction. Taken together, these findings reveal the emergence of a digitally mediated rural imaginary that is actively produced, circulated, and consumed on social media platforms. Although the topic of urban to rural migration in Romania is relatively narrow, and the three hashtags we analyze are in Romanian, the number of video views (24.3 million) has exceeded the total number of TikTok users in Romania (8.5 million) and Romanian speakers (22 million). This highlights that there is a form of digital rurality as an independent space in the Romanian context which is larger than would appear evident at first glance.

In what follows, we situate these findings within the framework of digital rurality, examining how the corpus reflects rural space across its three dimensions — digital rural localities, formal representations of the digital rural, and the everyday lives of the digital rural as introduced in \cite{HE2026103958}.

\subsection{Digital rural localities}


Global social media platforms are now available in remote areas across the globe, and the \textbf{socio-spatial changes} that follow from the proliferation of digital technologies in rural areas are multiple, by bringing forward \textit{new activities}, and producing \textit{new tensions} and \textit{new socio-spatial networks}. 

The urban to rural content creator migrants we studied, living in remote rural areas in Romania engage in \textbf{new activities}, such as streaming, but also remote digital work. In paid labor through social media platforms — those profiles that gather large followings can generate income through sponsorships from various companies (i.e. cosmetics, construction materials, small businesses, turistic venues) and political campaigns. This represents a significant shift in how rural economies function,  as it moves parts of the community from inward-facing, subsistence-oriented activities toward outward-facing, audience-dependent ones. Content about rural life — whether documenting renovation projects, local landscapes — can attract followers and, in turn, revenue, in areas where formal employment opportunities are scarce. These new forms of digital labour in part facilitate urban to rural migration. 

At the same time, this shift brings \textbf{new tensions}: creators who build large online audiences  alter the social fabric of communities that previously operated through more bounded, locally-embedded relationships, introducing new hierarchies of visibility and economic opportunity that do not necessarily align with existing community structures (this is even discussed to a certain extent in one of C3's videos). While it is clear that many of the creators spend significant amounts of time producing the social media content, within the corpus we analyze, we did not find any references to whether this becomes a source of income for them. Some of the creators we analyze express nationalist sentiment and advocate for Romanian sovereignty, in what appears to be sponsored political content, while simultaneously relying on platforms owned and governed by non-Romanian, corporate entities. This tension is most visible in creators such as C2, whose videos often open with appeals to "Romanian brothers," and C3, who frequently invokes the beauty of Romania and other times calls for wise leaders to resist foreign influence. The third hashtag itself — repopulamsateleromaniei ("we repopulate Romania's villages") — carries an implicit nationalist undertone. Whether or not creators are aware of this contradiction (the fact that these narratives of local rootedness and cultural sovereignty circulate on globally owned infrastructures) is not clear.  While it is evident that some creators invest considerable time and effort in producing their content, the videos rarely make explicit whether this labor is remunerated as none of the creators we analyzed openly disclosed brand partnerships or platform monetization arrangements in the videos we coded, despite some having follower counts — in the tens of thousands — that would plausibly place them within the micro-influencer range. This aligns with findings from broader TikTok research, which has documented blurring between personal narrative and promotional content \cite{Literat2026}. Taken together, these dynamics point to the emergence of new and ambiguous forms of labor in rural Romania. For individuals who have relocated to areas with limited formal employment opportunities, the platform may function as an economic lifeline as much as a space of self-expression. This is significant from a rural development perspective: the labor of representing rural life online is not captured in conventional economic statistics. At the same time, this labor is precarious and platform-dependent, subject to algorithmic changes and moderation policies over which creators have no control.

These \textbf{new socio-spatial networks} are, in this way connected globally. They also allow rural dwellers to be in conversations with others around the country, and around the world. At the same time, many of the videos in the corpus function as touristic material, promoting Romania to international audiences and contributing to informal country branding.

\subsection{Formal representations of the digital rural}


In the framework of digital rurality, formal representations of the rural refer to \textbf{how rural spaces are imagined and projected} by actors such as planners and goveranance institutions, developers, and digital technological corporations.

The most evident change over the last decade is the ways in which social media and other technological corporations complicate these imagined spaces. Data (which is often geo-located) produced for and owned by big tech is vast - and increases the informational density for places which were relatively remote and dis-connected. In the case of Romania, some of the rural areas remain poorly connected through physical infrastructure. Practically, this data comes in specific formats such as videos, images and comments to the audio-visual content. Creators now play direct roles in producing representations of the rural, generating a substantial body of digital representations about the Romanian countryside. These representations are diverse, as shown in Section 4.1. This is interesting given that rural Romania is frequently characterised in policy and academic literature as data-scarce \cite{Waschkova2024}. However, this content does not belong to local communities or Romanian authorities — it is hosted on, and governed by, platforms owned by large technology companies headquartered and owned outside Romania. The infrastructure through which rural Romania is being represented and made visible to the world is therefore not locally controlled, a tension that sits uneasily alongside the nationalist and sovereign sentiments that run through some of the content itself.

This complication is most evident in the areas of the corpus where political views and content are presented. Several creators, including C2 and C3, appear to produce content that mobilizes ideas of national pride, cultural heritage, religious orthodoxy and Romanian sovereignty — with some videos containing what appear to be sponsored political messages. The rural landscape itself becomes a backdrop for these claims: the need to preserve local traditions, protect cultural heritage, and resist foreign influence is invoked alongside images of cattle, mountain scenery, and traditional crafts. In this way, local and national politics are played out on digital platforms that monetize the production and distribution of political content and so local governance and in turn the leadership of institutions depend on these digital platforms. This became most evident in Romania's 2024 canceled election where one far right candidate rose to aprox. 20\% of votes in the first round based on an aggressive on-line campaign, especially on TikTok and where the hashtag of his name became the ninth most popular hashtag in the world a few days prior to the elections. The image that had been constructed around the candidate used rural landscapes, traditional costumes and lifestyles to reference environmental sustainability and build a narrative of national sovereignety and mobilize not the urban population, but those in rural areas. Nevertheless, this took many  living in the large urban centers, including academics by surprise \cite{Cistelecan02012025, Nistor2025}. 

\subsection{The everyday lives of the digital rural}

The everyday lives of the digital rural include \textbf{new identities and status} of urban to rural migrants, \textbf{new rural experiences} and \textbf{the virtual rural}.

The urban to rural migrants we analyze construct \textbf{new identities and status}: rurality is increasingly associated with living more sustainably, being more responsible towards nature, and caring for cultural and natural heritage. The urban to rural migrant identity is also constructed as someone with a higher degree of individual independence, with a can-do attitude, and better connected to their own needs (sometimes even framed as having a stronger connection to God), instead of being someone "working for the interests of others" (i.e. large corporations). \textbf{New rural experiences} involve on the one hand the fact that social media content makes rural landscapes available to broad audiences around the world (as seen for examples in the videos clustered under Nature\&Aestehtics in the topic modeling). Aspects of everyday life are also documented - such as caring for gardens and animals, cooking, or storing food. For the creators themselves, the time and effort spent on producing content represent new ways of engaging with the rural spaces. These together form a \textbf{virtual rural} where platform algorithmic logic aggregate scattered videos (snipped, carefully constructed representations) about rural life into a  narrative stream that might appear coherent, but remains representation. Features such as the comment section allow broad audiences to participate in constructing this virtual rurality, where they can engage with the creators.

\section{Conclusion}
This paper examined urban-to-rural migration in Romania through the lens of digital rurality, analysing a corpus of 901 TikTok videos posted under three Romanian-language hashtags that together accumulated 24.3 million views. Using both quantitative topic modelling and qualitative content analysis, we addressed three research questions concerning the motivations, profiles, and lived experiences of urban-to-rural movers as documented on TikTok.

With respect to motivations (RQ1), we find that urban-to-rural migration in Romania is driven by a combination of economic and lifestyle factors. Housing affordability in large urban centres such as Bucharest and Cluj pushes people toward rural areas, while the pull of slower living, self-sufficiency, and proximity to nature draws them in. The Covid-19 pandemic acted as a key enabler, normalising remote and flexible work arrangements that made rural relocation practically feasible for a segment of the urban population.

In terms of mover profiles (RQ2), the creators we analyse are predominantly young to middle-aged, often couples, including both domestic urban dwellers and members of the Romanian diaspora returning from Western Europe or North America. They tend to be digitally literate, with some already working as freelancers or digital nomads prior to their move.

With respect to lived experience (RQ3), the transition is presented on TikTok as largely positive, centred on self-sufficiency, identity reconstruction, and reconnection with nature and tradition. However, the platform's affordances encourage selective and aestheticised representations of rural life, with hardships rarely foregrounded.
Taken together, these findings illustrate the emergence of a digital rural public sphere in Romania, one that transcends the country's borders and reaches audiences well beyond the Romanian-speaking world. TikTok does not merely document this migration — it actively shapes how rural life is imagined, desired, and performed. This has implications for rural planners and policymakers, who may find in these platforms both a rich source of data on migration trends and a channel through which to reach and engage with new rural communities.
Future research could examine the gap between the romanticised rural life portrayed on TikTok and the actual experiences of movers, through interviews or longitudinal studies. The role of platform algorithms in amplifying particular narratives of rural life, and the tensions between nationalist sentiment and globally owned platforms, also warrant further investigation.


\bibliographystyle{elsarticle-harv} 
\bibliography{Urb-2-rur}

@misc{TikHub,
  author={TikHub},
  title={TikHub Douyin/ TikTok/ Xiaohongshu/ Bilibili/ Kuaishou/ Weibo/ Instagram/ YouTube/ Twitter/ Captcha Solver/Temp Mail API},
  howpublished={\url{https://docs.tikhub.io/}},
  year={2025}
}

@article{UNAYGAILHARD2023,
title = {An examination of digital empathy: When farmers speak for the climate through TikTok},
journal = {Journal of Rural Studies},
volume = {102},
pages = {103075},
year = {2023},
issn = {0743-0167},
doi = {https://doi.org/10.1016/j.jrurstud.2023.103075},

author = {İlkay Unay-Gailhard and Kati Lawson and Mark A. Brennan},
}

@article{HE2026103958,
title = {Digital rurality: A three-fold model and research agenda},
journal = {Journal of Rural Studies},
volume = {122},
pages = {103958},
year = {2026},
issn = {0743-0167},
doi = {https://doi.org/10.1016/j.jrurstud.2025.103958},
url = {https://www.sciencedirect.com/science/article/pii/S0743016725003997},
author = {Huiyan He and Agnieszka Leszczynski},
}

@article{HALFACREE2007125,
title = {Trial by space for a ‘radical rural’: Introducing alternative localities, representations and lives},
journal = {Journal of Rural Studies},
volume = {23},
number = {2},
pages = {125-141},
year = {2007},
issn = {0743-0167},
doi = {https://doi.org/10.1016/j.jrurstud.2006.10.002},
url = {https://www.sciencedirect.com/science/article/pii/S0743016706000696},
author = {Keith Halfacree},
}

@misc{Eurostat2026,
    key = {Eurostat},
    author = {Eurostat},
    title = {Key Figures on Europe},
    url = {https://ec.europa.eu/eurostat/cache/visualisations/keyfigures/},
    year = {2026},
    dateaccessed = {24.02.2026},

}

@misc{TikTok2026,
    key = {World Population Review},
    year = {2026},
    title = {TikTok users by country},
    dateaccessed = {24.02.2026},
    url = {https://worldpopulationreview.com/country-rankings/tiktok-users-by-country},
}

@article{Vais2022,
	author = {Dana Vais},
	title = {Systematization: A Key Term in 20th-Century Romanian Urbanism},
	journal = {Urban Planning},
	volume = {7},
	number = {1},
	year = {2022},
	issn = {2183-7635},	
    pages = {207--222},	
    doi = {10.17645/up.v7i1.4791},
	url = {https://www.cogitatiopress.com/urbanplanning/article/view/4791},
}

@article{HaoChen,
	author = {Han Chen and Min Wang and Zhen Zhang},
	title = {Research on Rural Landscape Preference Based on TikTok Short Video Content and User Comments},
	journal = {International Journal of Environmental Research and Public Health},
	volume = {19},
	number = {16},
	year = {2022},
    doi = { https://doi.org/10.3390/ijerph191610115},
	url = {https://www.mdpi.com/1660-4601/19/16/10115},
}

@CONFERENCE{Hutto2014,
  author  = {C.J. Hutto and Eric Gilbert},
  title   = {VADER: A Parsimonious Rule-based Model for Sentiment Analysis of Social Media Text},
  booktitle = {Proceedings of the Eighth International AAAI Conference on Weblogs and Social Media}, 
  year    = {2014},
  pages   = {216-225}
}

@article{Mulianingrum2025,
	author = {Restu Lestari Mulianingrum and Erwin Yudi Hidayat},
	title = {Comparative Performance of SVM and BERT-Base Using Hybrid Preprocessing for Fast Fashion Sentiment Analysis},
	journal = {Journal of Applied Informatics and Computing},
	volume = {9},
	number = {6},
	year = {2025},
    doi = {10.30871/jaic.v9i6.11385},
	url = {https://doi.org/10.30871/jaic.v9i6.11385},
}

@article{Hautea2021,
	author = {Samantha Hautea and Perry Parks and Bruno Takahashi and Jing Zeng},
	title = {Showing They Care (Or Don’t): Affective Publics and Ambivalent Climate Activism on TikTok},
	journal = {Sage Journals},
	volume = {7},
	number = {2},
	year = {2021},
    doi = {10.1177/20563051211012344},
}

@article{Hart2022,
	author = {Shana-Kay Hart},
	title = {Are You New Here? Understanding How Generation Z Uses TikTok and the Future of Social Media Marketing},
	journal = {University of MiamiProQuest Dissertations \& Theses},
	year = {2022},
    doi = {10.1177/20563051211012344},
	url = {https://www.sciencegate.app/document/10.1177/20563051211012344},
}

@article{Zha2026,
	author = {Mingyue Zha and Ho-Chun Herbert Chang},
	title = {Interpreting Multimodal Communication at Scale in Short-Form Video: Visual, Audio, and Textual Mental Health Discourse on TikTok},
	journal = {arXiv preprint},
	year = {2026},
	url = {https://arxiv.org/abs/2601.15278v1},
}

@article{Beach2025,
	author = {Sarah C. Beach and  Kirsten M. Bogunovich and  Samuel J. Borgemenke and  Sophia M. Thompson and  Allyson S. Hughes},
	title = {A Retrospective Analysis of Dermatologic Manifestations of Diabetes on TikTok},
	journal = {Clin Diabetes},
	volume = {43},
	number = {3},
	year = {2025},
    doi = {10.2337/cd24-0089},
	url = {https://doi.org/10.2337/cd24-0089},
}

@article{Zhang2025,
	author = {Yue Zhang and Suziana Mat Yasin and Li Sun},
	title = {The implications of urban-to-rural migration on rural development: A systematic literature review},
	journal = {GeoJournal},
	year = {2025},
    doi = {10.1007/s10708-025-11336-2},
	url = {https://doi.org/10.1007/s10708-025-11336-2},
}

@article{Erkan2025,
	author = {Öcek, Rüya Erkan and Tolga İslam},
	title = {Who Moves to the Countryside, and Why? A Systematic Review of Urban-to-rural Migration},
	journal = {Planlama},
        volume = {35},
	number = {1},
	year = {2025},
    doi = {10.14744/planlama.2024.00086},
	url = {https://10.14744/planlama.2024.00086},
}

@article{Carmen2022,
	author = {Delgado Viñas, Carmen and Gómez Moreno, María Luisa},
	title = {The interaction between urban and rural areas: an updated paradigmatic, methodological and bibliographic review},
	journal = {Land},
        volume = {11},
	number = {8},
	year = {2022},
    doi = {10.3390/land11081298},
	url = { https://doi.org/10.3390/land11081298},
}

@article{Li2025,
	author = {Jiewei Li and Dongxia Wei and Ming Lu},
	title = {Institution and Emotion— Examining Cross-Provincial Migration and Homesickness Expression through Video Big Data},
	journal = {Acedemic Monthly},
        volume = {3},
	year = {2025},
    doi = {DOI:10.12451/202503.02789},
	url = { https://pssxiv.cn/abs/202503.02789},
}

@article{Takahashi2021,
	author = {Yasuo Takahashi and 
Hiroyuki Kubota and Sawako Shigeto and  Takahiro Yoshida and Yoshiki Yamagata},
	title = {Diverse values of urban-to-rural migration: A case study of Hokuto City, Japan},
	journal = {Journal of Rural Studies},
        volume = {87},
	year = {2021},
    doi = {10.1016/j.jrurstud.2021.09.013},
	url = { https://doi.org/10.1016/j.jrurstud.2021.09.013},
}

@article{Jorge2025,
	author = {Vázquez-Herrero, Jorge and María-Cruz Negreira-Rey},
	title = {TikTok Journalism: A Methodological Approach to Content Analysis on the Short-Video Platform},
	journal = {Research Methods for Social Media Journalism},
	year = {2025},
}

@article{Tim2024,
	author = {Mavrič, Tim and Neža Čebron Lipovec},
	title = {Social Media Groups in Interaction With Contested Urban Narratives: The Case of Koper/Capodistria, Slovenia},
	journal = {Urban planning},
        volume = {9},
	year = {2024},
    doi = {10.17645/up.7083},
	url = {  https://doi.org/10.17645/up.7083},
}

@article{Brittany2025,
	author = {Quinn, Brittany and Lindsey Nichols and Jennifer Frazee and Mark Payton and Rachel MA Linger},
	title = {Dissemination of Information on Selective Serotonin Reuptake Inhibitors on TikTok: Analytical Mixed Methods Study of Creator Types, Content Tone, and User Engagement},
	journal = {JMIR Mental Health},
        volume = {12},
        number = {1},
	year = {2025},
    doi = {10.2196/77383},
	url = { https://doi.org/10.2196/77383},
}

@article{Lee2011,
	author = {Komito, Lee},
	title = {Social media and migration: Virtual community 2.0.},
	journal = {Journal of the American Society for Information Science and Technology},
        volume = {62},
        number = {6},
	year = {2011},
    doi = {10.1002/asi.21517},
	url = {  https://doi.org/10.1002/asi.21517},
}

@article{Rianne2014,
	author = {Dekker, Rianne and Godfried Engbersen},
	title = {How social media transform migrant networks and facilitate migration},
	journal = {Global networks},
        volume = {14},
        number = {4},
	year = {2014},
    doi = {10.1111/glob.12040},
	url = {   https://doi.org/10.1111/glob.12040},
}

@article{PETRESCU2022,
title = {From scythe to smartphone: Rural transformation in Romania evidenced by the perception of rural land and population},
journal = {Land Use Policy},
volume = {113},
pages = {105851},
year = {2022},
issn = {0264-8377},
doi = {https://doi.org/10.1016/j.landusepol.2021.105851},
url = {https://www.sciencedirect.com/science/article/pii/S0264837721005743},
author = {Ruxandra Malina Petrescu-Mag and Dacinia Crina Petrescu and Hossein Azadi},
}

@Article{Muntele2021,
AUTHOR = {Muntele, Ionel and Istrate, Marinela and Horea-Șerban, Raluca Ioana and Banica, Alexandru},
TITLE = {Demographic Resilience in the Rural Area of Romania. A Statistical-Territorial Approach of the Last Hundred Years},
JOURNAL = {Sustainability},
VOLUME = {13},
YEAR = {2021},
NUMBER = {19},
ARTICLE-NUMBER = {10902},
URL = {https://www.mdpi.com/2071-1050/13/19/10902},
ISSN = {2071-1050},
DOI = {10.3390/su131910902}
}

@article{ORSI2026,
title = {Street design and driving behavior: Evidence from a large-scale study in Milan and Amsterdam},
journal = {Transportation Research Part C: Emerging Technologies},
volume = {185},
pages = {105554},
year = {2026},
issn = {0968-090X},
doi = {https://doi.org/10.1016/j.trc.2026.105554},
url = {https://www.sciencedirect.com/science/article/pii/S0968090X26000422},
author = {Giacomo Orsi and Titus Venverloo and Andrea {La Grotteria} and Umberto Fugiglando and Fábio Duarte and Paolo Santi and Carlo Ratti},
}

@article{Hawelka2014,
author = {Bartosz Hawelka and Izabela Sitko and Euro Beinat and Stanislav Sobolevsky and Pavlos Kazakopoulos and Carlo Ratti},
title = {Geo-located Twitter as proxy for global mobility patterns},
journal = {Cartography and Geographic Information Science},
volume = {41},
number = {3},
pages = {260--271},
year = {2014},
publisher = {Taylor \& Francis},
doi = {10.1080/15230406.2014.890072},
URL = {https://doi.org/10.1080/15230406.2014.890072},

}

@article{Li2026,
    author = { Li, Maosu and Guo, Song and Duarte, Fábio and Kumar, Ashutosh and Kobori, Norimasa and Xue, Fan and Zhuang, Weimin and Yeh, Anthony G. O. and Ratti, Carlo},
    title = {Influence of objective and perceived exposures to urban nature on people’s happiness},
    journal = {npj Urban Sustainability},
    year = {2026},
    volume = {6},
    issue = {1},
    doi = {10.1038/s42949-025-00306-9}
}

@article{deSouza2025,
    author = {de Souza, Priyanka N and Shea, Amanda A and Vitzthum, Virginia J and Duarte, Fabio and Hanly, Claire Gorman and Timmons, Meghan and Huguelet, Patricia and Sammel, Mary D and Ratti, Carlo and Braun, Danielle and Nethery, Rachel C},
    title = {The effect of air pollution exposure on menstrual cycle health using self-reported data from a mobile health app: a prospective, observational study},
    journal = {The Lancet Planetary Health},
    publisher = {Elsevier},
    year = {2025},
    doi = {10.1016/S2542-5196(25)00080-4}
}

@article{KANG2023,
title = {Assessing differences in safety perceptions using GeoAI and survey across neighbourhoods in Stockholm, Sweden},
journal = {Landscape and Urban Planning},
volume = {236},
pages = {104768},
year = {2023},
issn = {0169-2046},
doi = {https://doi.org/10.1016/j.landurbplan.2023.104768},
url = {https://www.sciencedirect.com/science/article/pii/S0169204623000877},
author = {Yuhao Kang and Jonatan Abraham and Vania Ceccato and Fábio Duarte and Song Gao and Lukas Ljungqvist and Fan Zhang and Per Näsman and Carlo Ratti},

}

@article{deSouza2024,
    author = {de Souza, Priyanka and Anenberg, Susan and Makarewicz, Carrie and Shirgaokar, Manish and Duarte, Fabio and Ratti, Carlo and Durant, John L. and Kinney, Patrick L. and Niemeier, Deb},
    title = {Quantifying Disparities in Air Pollution Exposures across the United States Using Home and Work Addresses},
    journal = {Environmental Science \& Technology},
    year = {2024},
    doi = {https://doi.org/10.1021/acs.est.3c07926},
    volume = {58},
    issue = {1}
}

@article{Morra2024,
    author = {Morra, Diego and Zhu, Xiaosheng and Liu, Chang and Fu, Kyle and Duarte, Fábio and Mora, Simone and He, Zhengbing and Ratti, Carlo},
    title = {Mapping sidewalk accessibility with smartphone imagery and Visual AI: a participatory approach},
    journal = {Philosophical Transactions of the Royal Society A: Mathematical, Physical and Engineering Sciences},
    volume = {382},
    number = {2285},
    pages = {20240106},
    year = {2024},
    month = {11},
    issn = {1364-503X},
    doi = {10.1098/rsta.2024.0106},
    url = {https://doi.org/10.1098/rsta.2024.0106},
}

@article{Ceccato2026,
    author = {Ceccato, Vania and Kang, Yuhao and Abraham, Jonatan and Näsman, Per and Duarte, Fábio and Gao, Song and Ljungqvist, Lukas and Zhang, Fan and Ratti, Carlo},
    title = {What Makes a Place Safe? Assessing AI-Generated Safety Perception Scores Using Stockholm’s Street View Images},
    journal = {The British Journal of Criminology},
    volume = {66},
    number = {2},
    pages = {265-289},
    year = {2026},
    month = {03},
    issn = {0007-0955},
    doi = {10.1093/bjc/azaf017},
    url = {https://doi.org/10.1093/bjc/azaf017},

}

@misc{INS2026,
    title = {Populația după domiciliu la 1 ianuarie 2025},
    author = {Institutual Național de Statistică},
    year = {2026},
}

@article{Literat2026,
    author = {Ioana Literat, Constance De Saint Laurent and Vlad Glăveanu, Rhea Jaffer and Sonia Kim and Sophia Diplacido},
    title = {“It’s not about laziness, it’s about efficiency”: Youth Perspectives on Generative AI In Higher Education Through the Lens of TikTok},
    journal = {Digital Culture \& Education},
    year = {2026},
}

@article{Cistelecan02012025,
author = {Alex Cistelecan and Costi Rogozanu and Adina Marincea and Adrian Grama and Elena Trifan and Stefan Baghiu and Alexandra Mercescu and Cosmin Cercel},
title = {The global polycrisis and the Romanian elections of 2024},
journal = {Journal of Contemporary Central and Eastern Europe},
volume = {33},
number = {1},
pages = {167--179},
year = {2025},
publisher = {Routledge},
doi = {10.1080/25739638.2025.2482389},
URL = {https://doi.org/10.1080/25739638.2025.2482389},

}

@article{Waschkova2024,
    author = {Lenka Waschková Císařová and Iveta Jansová  and Jan Mota},
    title = {Delayed Reflections: Media and Journalism Data Deserts in the Post‐Socialist Czech Republic},
    journal = {Media and Communication},
    year = {2024},
    vol = {12},
    doi = {https://doi.org/10.17645/mac.7198}
}

@inbook{Horvath2024,
title = "Digital Twins in Architecture: An ecology of practices and understandings",
author = "Anca-Simona Horvath and Panagiota Pouliou",
year = "2024",
month = jan,
day = "1",
doi = "10.1201/9781003425724-46",
language = "English",
pages = "662--686",
editor = "Zhihan Lv",
booktitle = "Handbook of Digital Twins",
publisher = "CRC Press",
address = "Netherlands",
edition = "1",

}

@Article{Nistor2025,
AUTHOR = {Nistor, Andreea and Zadobrischi, Eduard},
TITLE = {The Virality of TikTok and New Media in Disrupting and Overturning the Election Cancellation Paradigm in Romania},
JOURNAL = {Administrative Sciences},
VOLUME = {15},
YEAR = {2025},
NUMBER = {11},
ARTICLE-NUMBER = {448},
URL = {https://www.mdpi.com/2076-3387/15/11/448},
ISSN = {2076-3387},
DOI = {10.3390/admsci15110448}
}

\end{document}